\documentclass[prb,amsmath,amssymb,twocolumn,showpacs,floatfix]{revtex4}
\usepackage{graphicx}
\usepackage[english]{babel}
\usepackage{times}
\usepackage{dcolumn}
\usepackage[usenames,dvipsnames]{color}
\usepackage{units}

\begin{document}

\title{The Kondo effect in the presence of the Rashba spin-orbit
interaction}

\author{Rok \v{Z}itko, Janez Bon\v{c}a}

\affiliation{Jo\v{z}ef Stefan Institute, Jamova 39, SI-1000 Ljubljana,
Slovenia,\\
Faculty  of Mathematics and Physics, University of Ljubljana,
Jadranska 19, SI-1000 Ljubljana, Slovenia}

\date{\today}

\begin{abstract}
We study the temperature scale of the Kondo screening of a magnetic
impurity which hybridizes with a two-dimensional electron gas in the
presence of the Rashba spin-orbit interaction. The problem is mapped
to an effective single-band impurity model with a hybridization
function having an inverse-square-root divergence at the bottom of the
band. We study the effect of this divergence on the Kondo screening.
The problem is solved numerically without further approximations using
the numerical renormalization group technique. We find that the Rashba
interaction leads to a small variation of the Kondo temperature
(increase or decrease) which depends on the values of the impurity
parameters.
\end{abstract}

\pacs{71.70.Ej, 72.10.Fk, 73.20.At}

\maketitle

\newcommand{\vc}[1]{\mathbf{#1}}

The spin-orbit (SO) interaction is a relativistic effect due to the
inter-dependence between electric and magnetic fields when considered from
different reference frames. It leads to a coupling between electron's
spin and its (orbital) motion in real space \cite{winkler}. The effect
is stronger in heavy elements from the bottom of the periodic system.
The spin-orbit interaction plays a central role in many proposals for
spintronic devices \cite{zutic2004} and it is the physical origin of
the recently discovered topologically non-trivial insulator phases
which do not break the time-reversal symmetry
\cite{kane2005f,bernevig2006sc,konig2007d,fu2007f}.
The spin-orbit interaction does not break the Kramers degeneracy, thus
the Kondo screening of magnetic moments is possible
\cite{meir1994,topo,feng2009} and should be observable in magnetic
adatoms on topological insulator surfaces or on thin layers of heavy
elements \cite{ast2007}.

The Kondo effect due to magnetic impurities in bulk simple metals was
shown to be strongly suppressed by doping the host material with small
ammounts of Pt impurities which have a large value of the SO coupling
constant \cite{gainon1969}, although these results have been
questioned and the expected trend with the increasing strength of the
the SO coupling has not been confirmed in later experiments
\cite{bujatti1976}. An Anderson impurity model with SO scattering term
for the conduction band electrons was studied by performing a
Schrieffer-Wolff transformation into an effective Kondo model
\cite{giovannini1971} and it captured some (but not all) of the
observations of Ref.~\onlinecite{gainon1969}. Kondo screening has also
been studied through weak-localization effects; the results indicate
that adding spin-orbit scatterers does not change the magnetic
scattering at all \cite{bergmann1986,meir1994}. It was pointed out
that the spin-orbit scatterers play the same role as elastic
nonmagnetic impurities because they do not break the time-reversal
symmetry \cite{meir1994}, thus the Kondo temperature is expected to
remain unchanged, in agreement with the experiment. For similar reason
of time-reversal invariance, the Kondo screening of magnetic moments
is expected to occur for magnetic impurities adsorbed on the surfaces
of three-dimensional topological insulators \cite{topo,feng2009},
unless the doping level is sufficiently high for the spontaneous
breaking of the time-reversal symmetry due to magnetic ordering. The
effect of the spin-orbit coupling in a two-dimensional (2D) electron
gas with Rashba interaction \cite{winkler} on the Kondo screening has
been studied in Ref.~\onlinecite{malecki2007}. The work was based on a
Kondo impurity Hamiltonian and after considering to which
conduction-band angular modes the impurity spin couples to, a
two-channel Kondo model was derived, in which each channel has a
different dispersion relation. By expanding the dispersion relations
to linear order, remarking that the Fermi velocity is the same in both
helicity channels, and after some further manipulations, the model
maps to a single-channel single-impurity Kondo model with unchanged
dimensionless Kondo coupling constant $\rho J$. In other words, it
was predicted that turning on the Rashba coupling (all other
parameters remaining the same) does not change the Kondo temperature
at all, except perhaps for some effective band-width effects. The same
problem has been considered starting from an Anderson impurity model
and using the Varma-Yafet variational ansatz \cite{feng2011}; this
work did not specifically address the question of the Kondo
temperature beyond noting that it remains constant in the limit of
small Rashba coupling, but focused instead on the spatial distribution
of the spin correlations. Recently, a new study of this system used a
Schrieffer-Wolff transformation to obtain an effective two-channel
Kondo model with an additional Dzyaloshinsky-Moriya (DM) interaction
\cite{zarea2011}. Using a scaling renormalization group analysis, it
was argued that the DM term renormalizes the Kondo coupling and
produces an exponential enhancement of the Kondo temperature. This
effect has not been observed in previous studies. Since all the cited
works are based on various mappings and approximations, it is
unclear which result for the variation of the Kondo temperature (or
lack thereof) is correct and what approximations are responsible for
the disagreement. For this reason, we reexamine the problem using an
approach where all the approximations are well controlled. To wit, we
integrate out the conduction-band degrees of freedom to obtain an
impurity action with a hybridization function which has an
inverse-square-root-divergence at the bottom of the conduction band.
This step is exact. We then numerically study the effective impurity
model using the numerical renormalization group (NRG) method
\cite{wilson1975,bulla2008}. While the NRG approach is not exact, the
approximations involved are all well controlled. We find that the
Kondo temperature is only weakly affected by the Rashba interaction:
it may weakly (approximately linearly) increase or decrease depending
on the values of the impurity parameters.

We study a magnetic impurity embedded in the 2D electron gas with
Rashba spin-orbit interaction
\cite{bychkov1984,winkler,malecki2007,feng2011,zarea2011}. The
impurity is described using the single-orbital Anderson impurity model
$H_0$ with an additional Rashba interaction term $H_\mathrm{SO}$:
\begin{eqnarray}
H_0 &=& \epsilon \left( n_\uparrow + n_\downarrow \right) + U
n_\uparrow n_\downarrow \\
 &+& \sum_{\vc{k}\sigma} \epsilon_k c^\dag_{\vc{k}\sigma}
c_{\vc{k}\sigma} + \sum_{\vc{k}\sigma} V_k \left( c^\dag_{\vc{k}\sigma} d_\sigma
+ \text{H.c.} \right), \\
H_\mathrm{SO} &=& \alpha \sum_\vc{k} \boldsymbol{\psi}^\dag_{\vc{k}} 
\left( k_x \boldsymbol{\sigma}_y - k_y \boldsymbol{\sigma}_x \right) 
\boldsymbol{\psi}_{\vc{k}} \\
&=& \alpha \sum_{\vc{k}} k e^{-i \phi_\vc{k}}
c^\dag_{\vc{k}\uparrow}
c_{\vc{k}\downarrow} + \text{H.c.}
\end{eqnarray}
The operator $d^\dag_\sigma$ creates an electron in the impurity
level, while $c^\dag_{\vc{k}\sigma}$ correspond to the conduction-band
electrons with the dispersion $\epsilon_k = k^2/2m^* + E_0$, where
$\vc{k}$ is the crystal momentum, $m^*$ the effective electron mass,
and $E_0$ the bottom of the conduction band. The occupancy operator is
$n_\sigma=d^\dag_\sigma d_\sigma$. The chemical potential is set to
the energy zero, $\mu=0$. In $H_\mathrm{SO}$, $\alpha$ parametrizes
the strength of the Rashba spin-orbit interaction,
$\boldsymbol{\psi}_{\vc{k}}$ is a spinor field $\left\{
c_{\vc{k}\uparrow}, c_{\vc{k}\downarrow} \right\}^T$, while
$\boldsymbol{\sigma}_{x/y}$ are Pauli matrices. $k$ and $\phi_\vc{k}$
are the polar coordinates of the wave number $\vc{k}$, with the polar
axis oriented along $k_y$.

We now switch to a continuum representation in a box of volume
$\mathcal{V}
\equiv 1$. The sums are transformed in the standard way as
$\sum_\vc{k} \to \mathcal{V}/(2\pi)^2 \int k\mathrm{d}k\mathrm{d}\phi$. The
operators are transformed as $c_{\vc{k}\sigma} \to (2\pi/\mathcal{V})
\psi^\dag_{\vc{k}\sigma}$; they are normalized such that $\{
\psi_{\vc{k}\sigma}, \psi^\dag_{\vc{k}'\sigma'} \}
=\delta^{(2)}(\vc{k}-\vc{k'})\delta_{\sigma\sigma'}$. Noting that
$\delta^{(2)}(\vc{k})=\delta(k)\delta(\phi)/k$, we switch to a
polar representation using $\psi_{\vc{k}\sigma} \to 1/\sqrt{k}
\xi_{k\phi\sigma}$, with $\{ \xi_{k\phi\sigma},
\xi^\dag_{k'\phi'\sigma'} \} = \delta(k-k') \delta(\phi-\phi')
\delta_{\sigma\sigma'}$. Finally, we switch to the angular momentum
basis $\xi_{k\phi\sigma} = \frac{1}{\sqrt{2\pi}} \sum_m e^{im\phi}
c^m_{k\sigma}$, where $m$ is the orbital magnetic quantum number
\cite{malecki2007,topo,zarea2011} and the anticommutation relations
take the form $\{ c^m_{k\sigma}, (c^{m'}_{k'\sigma'})^\dag \} =
\delta(k-k') \delta_{mm'} \delta_{\sigma\sigma'}$. Collectively, these
transformation steps can be written as
\begin{equation}
c_{\vc{k}\sigma} \to \sqrt{\frac{2\pi}{k}} 
\sum_{m=-\infty}^{\infty} e^{im\phi} c^m_{k\sigma}.
\end{equation}
We then introduce the linear combination which diagonalizes the
conduction-band Hamiltonian including the Rashba coupling terms
\cite{malecki2007,zarea2011}
\begin{equation}
c^{m+1/2}_{kh} = \left( c^{m}_{k\uparrow}
+ h c^{m+1}_{k\downarrow} \right)/\sqrt{2},
\end{equation}
where $h = \pm 1$ is the chirality quantum number. The corresponding
band energies are
\begin{equation}
\begin{split}
\epsilon_{kh} = \epsilon_k + \alpha k h = 
\frac{(k + h k_0)^2}{2m^*} + E_0 - E_R,
\end{split}
\end{equation}
where $k_0=m^* \alpha$ is the Rashba momentum, and $E_R = k_0^2/2m^*$
is the Rashba energy. This implies that the bottom of the conduction
band shifts from $E_0$ to $E_0-E_R$ upon switching on the Rashba
interaction.

\begin{figure}[htbp]
\centering 
\includegraphics[clip,width=8cm]{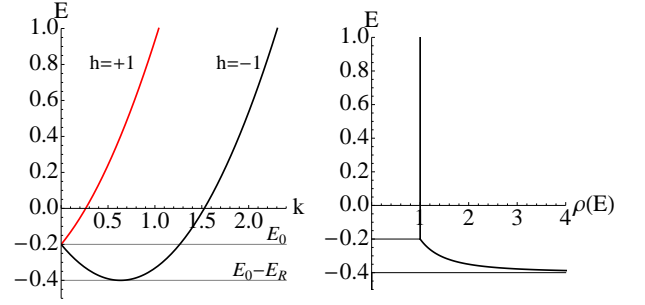}
\caption{(Color
online) The dispersion and the density of states in the presence of
the Rashba spin-orbit coupling. The label $h=\pm1$ indicates the
helicity of the band.} \label{fig1}
\end{figure}

We obtain the effective impurity model
\begin{equation}
\begin{split}
H &= \sum_{hm} \int_0^{\infty} \mathrm{d}k \epsilon_{kh} \left( c_{kh}^{m+1/2} \right)^\dag
c_{kh}^{m+1/2} \\
&+ \epsilon \left( n_\uparrow + n_\downarrow
\right) + U n_\uparrow n_\downarrow \\
&+ \frac{1}{2\pi} \sum_{h} \int_0^{\infty} \sqrt{k} \mathrm{d}k V_k
\frac{\sqrt{2\pi}}{\sqrt{2}}  \times \\
& 
\quad \left[
\left( c_{kh}^{1/2} \right)^\dag d_\uparrow 
+ (-1)^{\frac{h-1}{2}} \left( c_{kh}^{-1/2} \right)^\dag d_\downarrow +
\text{H.c.}
\right].
\end{split}
\end{equation}
We now integrate out the conduction-band modes and obtain an
effective impurity action of the following form:
\begin{equation}
\begin{split}
S &= \int_0^\beta \mathrm{d}\tau \left[ \sum_\sigma d^\dag_{\sigma} \left(
\frac{\partial}{\partial\tau} + \epsilon \right) d_\sigma
+ U n_\uparrow n_\downarrow \right] \\
&+ \frac{1}{\pi} \sum_h \int_0^\beta \mathrm{d}\tau \int_0^\beta
\mathrm{d}\tau' d^\dag_{\sigma} \Delta_h(\tau-\tau') d_{\sigma}.
\end{split}
\label{S}
\end{equation}
The hybridization function $\Delta_h(\tau)$ is the Fourier transform
of
\begin{equation}
\Delta_h(i\omega_n) = \frac{1}{4\pi} \int_0^\infty k \mathrm{d}k
\frac{V_k^2}{i\omega_n - \epsilon_{kh}}.
\end{equation}
While $\Delta_+(i\omega_n)$ and $\Delta_-(i\omega_n)$ are somewhat
complicated functions, their sum $\Delta=\sum_h \Delta_h$ is simple.
Assuming $V_k \equiv V$, the imaginary part $\Gamma(E)=-1/(2\pi)
\mathrm{Im} \Delta(E+i0^+)$ is found to be of the same functional form as
the density of states (DOS) of the conduction band:
\begin{equation}
\Gamma(E) = \Gamma_0 \left\{ 
\begin{array}{ll}
0 & E<E_0-E_R \\
\sqrt{\frac{E_R}{E-(E_0-E_R)}} & E_0-E_R < E < E_0 \\
1 & E>E_0
\end{array}\right.
\end{equation}
where $\Gamma_0=\pi \frac{m^*}{2\pi} V^2$. This result is remarkably
simple and could have been guessed in advance, since the impurity is
assumed to be point-like and simply couples to the local DOS at its
position. The only effect of the Rashba interaction as far as the
local impurity properties are concerned is the emergence of an
additional energy interval $[E_0-E_R;E_0]$ with finite density of
states which diverges with an inverse-square-root divergence at the
lower limit, see Fig.~\ref{fig1}. The effect of this diverging density
of states is difficult to evaluate analytically, therefore we resort
to using a numerical technique.

We compute the impurity spectral function using the numerical
renormalization group (NRG) method
\cite{wilson1975,bulla2008} with extensions
for arbitrary density of states
\cite{oliveira1994,chen1995,ingersent1996,bulla1997,vojta2002eurpjb,campo2005,resolution,odesolv}.
NRG can handle diverging DOS both at the Fermi level 
\cite{vojta2002eurpjb,vanhove}, as well as away from the Fermi level
\cite{peters2008nnn}. Here the discretization has been performed using
the discretization scheme from Refs.~\onlinecite{resolution,odesolv};
this scheme easily handles inverse square root divergencies at finite
frequencies without a need for introducing artificial cut-offs. NRG
parameters were $\Lambda=2$, twist-averaging over $N_z=64$
discretization grids, and the truncation cutoff set at $10\omega_N$,
where $\omega_N$ is the characteristic energy scale at the $N$-th
NRG step.

\begin{figure}[htbp]
\centering 
\includegraphics[clip,width=8cm]{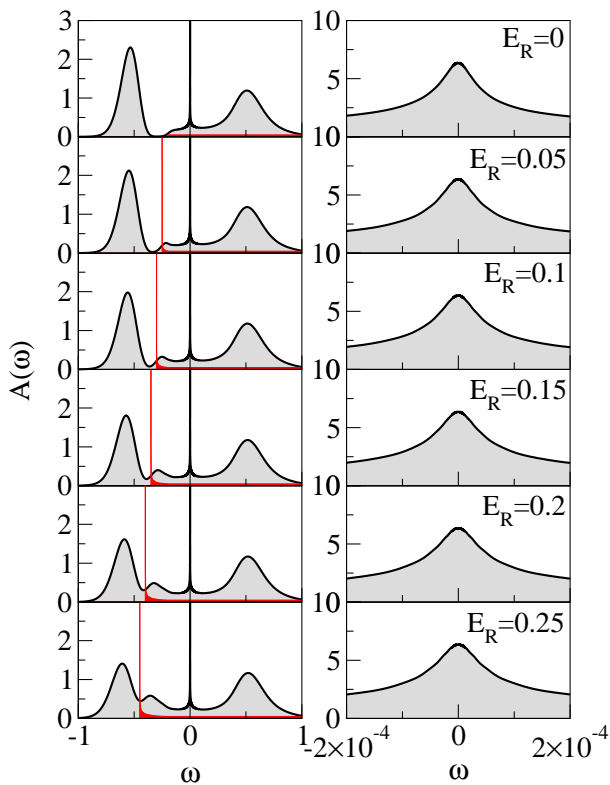}
\caption{(Color online) The impurity spectral functions for a range of
the Rashba spin-orbit coupling strengths. The gray curve (red online)
in the left-hand panels is the effective hybridization function that
the impurity state mixes with. The parameters are $U=1$,
$\epsilon=-0.5$, and the hybridization strength at Fermi level for
$E_R=0$ has been set to $\Gamma_0=0.05$. The band bottom is $E_0=-0.2$.} \label{fig2}
\end{figure}

\begin{figure}[htbp]
\centering
\includegraphics[clip,width=8cm]{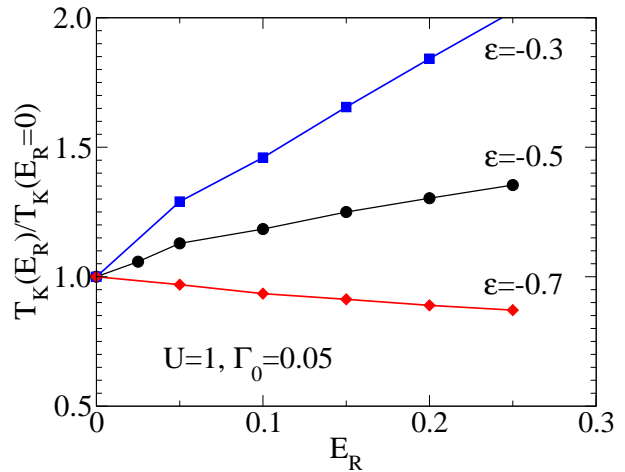}
\caption{(Color online) The Kondo temperature calculated using the
   NRG. The parameters are specified in the plot.
   The temperatures are rescaled by the Kondo temperature in
   the absence of the Rashba interaction.
}
\label{fig3}
\end{figure}

The calculated spectral functions are shown in Fig.~\ref{fig2}. Left
panels show the spectral function over the full energy interval, while
the right panels are a close-up on the Kondo resonance at the Fermi
level. The hybridization function entering the impurity model is shown
as a gray curve (red online) in the left-hand panels. The spectral
function features two atomic resonances, one at $\omega=\epsilon$
(corresponding to the extraction of an electron) and one at
$\omega=\epsilon+U$ (corresponding to the addition of an electron).
The width of atomic spectral peaks is approximately $2\Gamma(\omega)$,
where $\omega$ is the peak position \cite{pruschke1989,logan1998}. The
peak at $\omega=\epsilon$ lies outside the band for all values of
$E_R$ considered, thus it should have zero width (i.e., it is a delta
peak); the finite width in the presented spectral functions is due to
the spectral broadening in the NRG. In addition to the atomic peaks,
the spectral function features the many-body Kondo resonance which
peaks near the Fermi level. The long logarithmic tails of the Kondo
resonance are asymmetric due to the particular energy dependence of
the hybridization function. At the bottom of the band, at
$\omega=E_0-E_R$, the spectral function has a maximum, then it drops
to zero at lower frequencies since in that range there are no states
in the band. Again, the drop to zero in presented spectral functions
is overbroadened due to technical reasons. 

The Kondo temperature (defined as in Ref.~\onlinecite{krishna1980a})
has been determined from the impurity magnetic susceptibility, but it
could alternatively be extracted from the Kondo resonance peak widths.
The results for a several different choices of the impurity parameters
are shown in Fig.~\ref{fig3}. Note that we have included parameter sets
both with $\epsilon+U/2=0$ and with $\epsilon+U/2 \neq 0$. We find
that the Kondo temperature exhibits some variation as a function of
$E_R$, however the variation is not exponential \cite{zarea2011}, but
rather linear in the large-$E_R$ limit with more complex variation for
small values of $E_R$ (which can be attributed to the emergence of the
inverse-square-root divergence in the hybridization function).
Depending on the value of $\epsilon$, $U$, and $\Gamma_0$, the Kondo
temperature is either an increasing or a decreasing function of
$E_R$. 

In systems with constant hybridization function $\Gamma$, the
Kondo temperature is given approximately by
\begin{equation}
\label{tk}
T_K = {\tilde D} \sqrt{\rho J} \exp\left(- \frac{1}{\rho J} \right),
\end{equation}
where $\rho J = 8\Gamma/\pi U$ is the dimensionless Kondo coupling
constant. If the hybridization function is energy-dependent, we have
to use its value at the Fermi level ($\Gamma_0$) in the expression
above. In the presence of the Rashba interaction, if the bottom of the
conduction band is initially below the Fermi level (i.e., $E_0<0$),
the density of states at $\omega=0$ will remain unchanged when the
Rashba coupling is increased. This explains the absence of an exponential
renormalization of $T_K$ and is in agreement with 
Ref.~\onlinecite{malecki2007}. Nevertheless, in Eq.~\eqref{tk} the
effective bandwidth ${\tilde D}$ does depend on the details of the
hybridization function, thus some variation of $T_K$ is in fact
expected and indeed observed, as shown in Fig.~\ref{fig3}.

We note that the effective impurity action, Eq.~\eqref{S}, has the
form of a single-band problem. In fact, the single-orbital Anderson
impurity model is always effectively a single-band problem, no matter
what kind of conduction-band it couples to \cite{bulla1997,topo}. The
required unique effective conduction band can always be constructed by
taking the combination of states which couples to the impurity orbital
(here a state proportional to the linear combination $\sum_\vc{k} V_k
c^\dag_{\vc{k}\sigma}$) as an initial state in the Gram-Schmidt
orthogonalization procedure applied on the conduction-band
Hamiltonian. Consequently, one obtains a single-band tight-binding
Hamiltonian, while all other conduction-band states are fully
decoupled from the impurity. This argument holds universally. 
A multi-channel Kondo model can be in some cases obtained from
the single-orbital Anderson impurity model by using
a non-optimal basis for the conduction-band modes that also
includes states which are in reality fully decoupled from the impurity.
These states
should not in any way affect impurity properties. Unfortunately, the
complete irrelevance of the redundant conduction-band states can only
be observed if the problem is solved exactly. If the problem is,
however, approached with an approximate method, there exist a
potential pitfall: the presence of the redundant states can affect the
impurity properties in some spurious way.

We also note that the problem of an impurity in the two-dimensional
electron gas can never be truly particle-hole symmetric, because the
conduction-band itself is not particle-hole symmetric. We therefore do
not expect any particular difference in the role of the Rashba
spin-orbit interaction depending on whether $\epsilon+U/2$ is zero or
not \cite{zarea2011}.

The predictions of this work could be experimentally tested in a
system where the spin-orbit interaction can be tuned, such as 2D
electron systems in semiconductors with metal gates. Furthermore, the
rapid advances in the field of ultra-cold atom system suggest that it
might become possible to build 2D fermionic gases with tunable SO
interaction \cite{liu2009so,dalibard2010,lin2011,chapman2011} and
couple it to a magnetic impurity \cite{recati2005}.

\begin{acknowledgments}
Discussions with Anton Ram\v{s}ak and Toma\v{z} Rejec and the support
of the Slovenian Research Agency (ARRS) under Program P1-0044 are
acknowledged.
\end{acknowledgments}

\bibliography{rashba}

\end{document}